\documentstyle[12pt]{article}
\title{Chaotic dynamics of high order \\ neural networks\thanks{E-
mails:  {\tt lemke@if1.ufrgs.br}, {\tt arenzon@if1.ufrgs.br} and
{\tt tamarit@brlncc.bitnet}}}
\author{Ney Lemke and Jeferson J. Arenzon  \\
Instituto de F\'{\i}sica \\
Universidade Federal do Rio Grande do Sul  \\ C.P. 15051 --
91501-970 -- Porto Alegre -- RS -- Brazil \\
\mbox{} \\
Francisco A. Tamarit \\
Centro Brasileiro de Pesquisas F\'{\i}sicas \\
Rua Xavier Sigaud, 150 \\
22290 -- Rio de Janeiro -- RJ -- Brazil}
\date{December 28, 1993}

\begin{document}
\thispagestyle{empty}
\maketitle\vspace{-1cm}

\begin{abstract}
The dynamics of an extremely diluted neural network with high order
synapses acting as corrections to the
Hopfield model is investigated. As in the fully connected case, the
high order terms  may strongly improve the storage capacity of the
system. The dynamics displays a very rich behavior, and in
particular a new chaotic phase emerges depending on the weight of
the high
order connections $\varepsilon$, the noise level $T$ and the
network load defined as the rate between the number of stored
patterns and the mean
connectivity per neuron $\alpha =P/C$.
\end{abstract}

\newpage

\section{INTRODUCTION}

Neural networks have been subject of intense research in
Statistical Mechanics in the last decade. However, besides
the equilibrium properties that are reasonable well understood in
the framework of spin glass theory, dynamical properties are hard
to treat analytically and still requires further work. In the limit
of low loading or focusing only in the first few time steps,
several treatments \cite{cool} yielded important results, the
challenge remaining for saturated networks \cite{russos}.
Particularly, for extremely and asymmetrically diluted networks the
dynamics can be exactly solved \cite{derrida}. It is worth to
stress here that both dilution and asymmetry are biological
realistic ingredients absent in the fully connected Hopfield model.

In order to introduce asymmetry in the
connections, the synapses $J_{ij}$ and $J_{ji}$  are cut  with
probability $1-C/N$  independently of each other \cite{derrida},
where $C$ is the mean connectivity of each neuron and the extreme
dilution limit is obtained when $C \ll \ln N$ ($N$ is the size of
the network).

Measuring the network load by $\alpha=P/C$, where $P$ is the number
of stored patterns,
Derrida, Gardner and Zippelius (hereafter DGZ) found that
there is a critical value $\alpha_c=2/\pi$ above which the system
cannot retrieve the stored information. Since the  storage capacity
is here measured by the amount of bits that can be stored ($PN$)
per synapse ($CN$), in the diluted version the Hopfield model is
more efficient than  the
fully connected case ($C=N$) in which $\alpha_c\simeq 0.138$
\cite{amit}.

Synapses connecting more than two neurons can be considered to both
improve the storage capacity
of the network \cite{per}-\cite{trs} and to mimic
real synapses existing in nervous systems  (see
\cite{per} and references therein).
For instance, axon-axon-dendrite connections
can be described as third order synapses (even more
intricate  connections, involving more than two axons, may also
exist in the brain) \cite{per}. Moreover, when some pairwise
connections are  close enough, they may interact somehow, and  that
can also be considered as high (even) order synapses. Diluted
networks with high order connections were studied in refs.
\cite{kanter,tama,wang}. Kanter \cite{kanter} and Tamarit {\it et
al} \cite{tama} considered a monomial of
order $k$ in the overlap and found a discontinuous transition at
$\alpha_c(k)$ for $k>2$. Also, in the $(\alpha,T)$ phase diagram
the retrieval phase shrinks as $k$ increases: $\alpha_c(k)\to 0$ as
$k\to\infty$ although the retrieval is perfect ($m\to 1$).   Wang
and Ross \cite{wang} treated instead a polynomial, controlling  the
relative coefficients (weights) and
found, besides a retrieval phase, regions where the system can be
either periodic or chaotic, depending on the noise or the degree of
parallelism in the updating.

In this paper we study the dynamical effects of introducing  a
correction term to the Hopfield model \cite{trs} in the extreme
dilution limit of both, two and fourth order synapses. The
difference between our model and those studied by Kanter {\em et
al} \cite{kanter}, Tamarit  {\em et al} \cite{tama} and
Wang {\em et al} \cite{wang} lays in the nature of high order
connections we consider (see details below).
The paper is organized as follows: section \ref{trunc.sec} reviews
the model and in sections \ref{dilu.sec}  and \ref{results.sec}
analytical and numerical results are presented
for the dynamics. In section \ref{conc.sec} we summarize and
present  our conclusions.

\section{THE TRUNCATED  MODEL}
\label{trunc.sec}

In a binary neural network, as for example
the Hopfield model\cite{hop,model},  each neuron is modeled by an
Ising variable $S_i$ that can take the values $\{+1,-1\}$
representing the active and
passive state, respectively.
Possible states of the network are given by $N$-dimensional vectors
$\vec{S}\in \{ +1,-1 \}^N$ and the embedded memories are
associated to  $P$  of these states,  denoted by $\vec{\xi}^{\mu}$
($\mu=1,\ldots,P$).

The fully connected  Truncated model can be regarded as the
Hopfield model plus correction terms. Since the synapses are
symmetric, the dynamics is ruled by a
Lyapunov function (called Hamiltonian by analogy with magnetic
systems) and the long time behavior of the network can be inferred
from a thermodynamical analysis of the system. In this case the
Hamiltonian is given by \cite{rsana,trs}
\begin{eqnarray}
\label{hamil.eq}
{\cal H}&=&\frac{N}{2} \sum_{\ell=1}^M (-1)^{\ell}
\varepsilon_{2\ell}  \sum_{\mu_1 < \ldots < \mu_{\ell}} m_{\mu_1}^2
m_{\mu_2}^2 \ldots m_{\mu_\ell}^2 \nonumber \\ &=&  -
\varepsilon_2 \frac{N}{2} \sum_{\mu_1} m_{\mu_1}^2 +  \varepsilon_4
\frac{N}{2}
\sum_{\mu_1<\mu_2} m_{\mu_1}^2 m_{\mu_2}^2 - \ldots + (-1)^M
\varepsilon_{2M} \frac{N}{2} \sum_{\mu_1 < \ldots <
\mu_M}  m_{\mu_1}^2 \ldots m_{\mu_M}^2
\label{tobetruncated}
\end{eqnarray}
where $2M$ is the highest order of interaction and
the overlap $m_{\mu}$ between the state of the
network $\vec{S}$ and the pattern $\vec{\xi}^{\mu}$ is given by
\begin{equation}
m_{\mu}=\frac{1}{N} \sum_{i=1}^N \xi_i^{\mu} S_i \:\:\:\:\:.
\end{equation}
In what follows we consider only the first correction
to the Hopfield term ($M=2$ and
$\varepsilon_{2\ell}~=~\delta_{1\ell}  + \varepsilon
\delta_{2\ell}$). Eq.(\ref{tobetruncated}) can then be rewritten as
\begin{equation}
{\cal H}=-\frac{1}{2} \sum_{i,j} J_{ij} S_i S_j +
\frac{\varepsilon}{2} \sum_{i,j,k,l} J_{ijkl} S_i S_j S_k S_l
\;\;\;\; .
\label{liapunov}
\end{equation}
The learning rule for the second order $J_{ij}$
and fourth order $J_{ijkl}$ follows directly from
eq.(\ref{tobetruncated})  and will be detailed in the next section
\cite{model}.

The equilibrium statistical mechanics \cite{amit} of this model was
studied in refs. \cite{trs}, showing an extremely rich
behavior that depends on the value of $\varepsilon$. The main
features of the system at $T=0$ (for non zero
temperatures the behavior is analogous) are summarized
by the phase diagram in fig.\ref{fcpdt0}: for large $\varepsilon$
($\sim 0.5$) the  overlap $m$ decreases continuously with $\alpha$,
going monotonically from 1 down to zero at
$\alpha_c^+(\varepsilon)$; as $\varepsilon$
decreases ($\sim$ 0.36),  the overlap presents a local minimum,
before finally going to zero at $\alpha_c^+ (\varepsilon)$. If
$\varepsilon<\varepsilon_c\simeq 0.3587$, the minimum yields a gap
separating two retrieval regions. Not too close to
$\varepsilon_c$ the gap $\Delta$ has a discontinuous border  at
$\alpha_c'$ and a continuous one at $\alpha_c^-$. The points where
$m$ continuously  approaches zero, $\alpha_c^{\pm}(\varepsilon)$,
are
\begin{equation}
\alpha_c^{\pm}(\varepsilon)=\left(\frac{1}{\sqrt{\varepsilon}} \pm
\sqrt{\frac{2}{\pi}}  \:\right)^2
\:\:\:\: .
\label{alfac}
\end{equation}
As $\varepsilon\to 0 $, $\Delta\equiv \alpha_c^-(\varepsilon) -
\alpha_c'(\varepsilon)$ goes to
infinity as $\varepsilon^{-1}$. Thus, when the Truncated model
recovers the Hopfield model ($\varepsilon\to 0$), the location of
the second retrieval region in the $\alpha$-axis goes to infinity
and $\alpha_c'(0)\simeq 0.138$. Numerical simulations \cite{trs} at
$T=0$ show the above transitions and the existence of the gap. The
basins of attraction seem  to be large
and $\alpha$-independent; the mean convergence time increases with
$\alpha$ and, for non small values of $\alpha$,
does not depend on the initial overlap.

There is also an optimal value of $\varepsilon$ that strongly
improves the storage capacity of the model by canceling the noise
term for each value of $\alpha$ \cite{trs}:
\begin{equation}
\varepsilon_{opt} = \frac{1}{1+\alpha} \;\;\;\; .
\label{epsopt}
\end{equation}
As we will see in the next two sections, there is an analogue of
this optimal value for the diluted network. A complete discussion
of these and other results for the fully connected Truncated model
can be found in refs.\cite{trs}.

\section{THE DILUTED NETWORK}
\label{dilu.sec}

In this section we study the dynamics of an extremely diluted and
asymmetric
version of the  network described in section \ref{trunc.sec} in two
different cases: the initial state has a macroscopic overlap with
a) only one
pattern and b) two correlated patterns (out of $P-2$ uncorrelated
ones). Due to the asymmetry in the connections, a Lyapunov function
can no longer be defined for the network and hence we are
constrained  to study the time evolution  of the system ruled by
the Heat Bath dynamics, given by
\begin{eqnarray}
S_i(t+\Delta t) = \left\{
\begin{array}{l}
+1 \mbox{   with probability   } (1+\exp [-2\beta_o h_i(t)] )^{-1}
\\ \mbox{} \\
-1 \mbox{   with probability   } (1+\exp [+2\beta_o h_i(t)] )^{-1}
\end{array}
\right.
\end{eqnarray}
where the parameter $\beta_o\equiv T_o^{-1}$ (called the inverse of
the temperature) measures the noise
level of the net and
$h_i(t)$ is the local field acting on the neuron $i$ at time $t$:
\begin{equation}
h_i=\sum_j J_{ij}S_j -\varepsilon\sum_{j,k,l}J^\prime_{ijkl}S_j S_k
S_l \;\;\;\; .
\end{equation}
Based on eq.(\ref{hamil.eq}), the couplings are:
\begin{eqnarray}
J_{ij}&=&C_{ij}\sum_\mu \xi^\mu_i\xi^\mu_j \\
J^\prime_{ijkl}&=&\frac{1}{3}\left(
J_{ijkl}+J_{ljki}+J_{kjil}\right) \\
J_{ijkl}&=&C_{ijkl}\sum_{\mu\neq\nu}
\xi^\mu_i\xi^\mu_j\xi^\nu_k\xi^\nu_l
\end{eqnarray}
and $C_{ij}$ and $C_{ijkl}$ are random variables distributed
according to the probabilities $\rho$ and $\tilde{\rho}$:
\begin{eqnarray}
\rho (C_{ij}) &=& \frac{C}{N}\delta (C_{ij}-1)+ \left( 1-
\frac{C}{N}\right)  \delta (C_{ij}) \nonumber \\ \mbox{} \\
\tilde{\rho} (C_{ijkl})&=&\frac{C}{N^3}\delta (C_{ijkl}-1) +
\left(1-\frac{C}{N^3}\right) \delta (C_{ijkl}) \nonumber
\end{eqnarray}
where $C \ll \log N$. The asymmetry is introduced
through the independence of $C_{ij}$ and $C_{ji}$ (the same holds
for $C_{ijkl}$).

In the fully connected case some of the self-connections are not
zero, particularly, $J_{iikl}$ and $J_{ijkk}$.
These self-interactions create correlations between the states of
neuron $i$ at different times, that is,
$\left\langle S_i(t) S_i(t^{\prime}) \right\rangle \neq 0$, what
apparently prevents from using the DGZ prescription to obtain the
time evolution of the diluted network. Actually, it can be proved
that the field generated by the self-interactions vanishes in the
thermodynamical limit, implying null correlations.

Assuming that the updating is parallel and the initial state is
correlated  with only one of
the embedded memories, we are interested in obtaining a recurrent
equation, $$
m(t+1) = f(m(t)) \;\;\;,
$$
for the overlap $m(t)$ between the state of the network and this
memory,
\begin{equation}
m(t)=\frac{1}{N}\sum_i \langle \xi_i^1 S_i (t) \rangle \;\;\; ,
\end{equation}
where $\langle \ldots \rangle $ denotes both thermal and
configuration averages. Using standard techniques and considering
non biased patterns, after taking the limit of $C\rightarrow
\infty$, the equation ruling the parallel evolution of the system
reads:
\begin{equation}
m(t+1)=\int_{-\infty}^{+\infty} {\cal D}y\;\tanh\,\beta\left[ m(t)-
\sqrt{2\alpha}y(1-\varepsilon m^2(t)) \right] \;\;\;\; ,
\label{mapm}
\end{equation}
where $C\alpha=P-1$, $\beta=C/T_o$ and ${\cal D}y$ is the gaussian
measure:
\begin{equation}
{\cal D}y = \frac{dy}{\sqrt{\pi}} e^{-y^2} \;\;\;\; .
\end{equation}

We also study the case where the initial state has a macroscopic
overlap only with the two first memories, which have a
fixed overlap between them,
\begin{equation}
\kappa =\frac{1}{N}\sum_i \xi_i^1\xi_i^2 \;\;\;\; ,
\end{equation}
the remaining $P-2$ being uncorrelated.
The quantities of interest are then the overlaps with
these two memories:
\begin{eqnarray}
m_1(t) &=& \frac{1}{N}\sum_i \langle \xi_i^1 S_i (t)\rangle
\nonumber  \\ \mbox{}  \\
m_2(t) &=& \frac{1}{N}\sum_i \langle \xi_i^2 S_i (t)\rangle
\nonumber \end{eqnarray}
and the time evolution equations for them are ruled by:
\begin{eqnarray}
M(t+1) &=& (1+\kappa)\int {\cal D}y \,
\tanh\beta\left[M\left(1-\varepsilon\frac{(M^2-
m^2)}{4}\right) -y\sqrt{2\alpha}\left(1-\frac{\varepsilon
(M^2+m^2)}{2}\right)\right] \nonumber \\ \mbox{} \nonumber \\
\mbox{} \\ \mbox{} \nonumber \\
m(t+1) &=& (1-\kappa)\int {\cal D}y \,
\tanh\beta \left[m\left(1+\varepsilon\frac{(M^2-
m^2)}{4}\right) -y\sqrt{2\alpha}\left(1-\frac{\varepsilon
(M^2+m^2)}{2}\right)\right] \nonumber
\label{dupla}
\end{eqnarray}
where:
\begin{eqnarray}
M &=& m_1+m_2 \nonumber \\ \mbox{} \\
m &=& m_1-m_2 \;\;\;\; . \nonumber
\end{eqnarray}

In the next section the above equations are numerically solved and
the results with their interpretation are presented.

\section{RESULTS}
\label{results.sec}

In the $T=0$ limit, $f(m)$ becomes
\begin{equation}
f(m)= \mbox{erf} \left[ \frac{m(t)}{\sqrt{2\alpha}(1-\varepsilon
m^2(t) )}  \right] \;\;\;\; .
\label{fmt0}
\end{equation}
Fig.\ref{mapa} displays this map for two different values of
$\varepsilon$ showing that for $\varepsilon<1$ it is
continuous while it is discontinuous otherwise.

In fig.\ref{meps} we show the fixed points of
the equation $m(t+1)=f(m(t))$ as a function of $\alpha$ for several
values of $\varepsilon<1$.
For $\varepsilon < 0.266$ the system has only a fixed point that
decreases continuously down to 0 at $\alpha_c = 2/\pi$. Above this
critical value, there is only the $m=0$ solution. For
$0.266 <\varepsilon< 1$ the transition is discontinuous and
$\alpha_c$ is greater than $2/\pi$, increasing with $\varepsilon$.
In this case there are three different regimes: for $\alpha<2/\pi$
there is only the retrieval solution; for $2/\pi < \alpha <
\alpha_c$ this solution coexist with the $m=0$ one and for
$\alpha>\alpha_c$ there is only the paramagnetic solution. In the
intermediate regime, an unstable solution appears that separates
the basins of attractions of the  two stable solutions and the
retrieval basin decreases with $\alpha$ as can be seen in
fig.\ref{meps}.
Also, for a given value of $\alpha$, the closer $\varepsilon$ is to
1 the better is the retrieval. The retrieval region below the line
$\alpha_c$, where exists a nonzero fixed point, is named R$_1$,
while above $\alpha_c$ the paramagnetic phase is named $P$.

At $\varepsilon=1$ there always is
the solution $m=1$ and the unstable solution found for
$\alpha>2/\pi$ drops to zero as we approach $\varepsilon=1$. Hence,
the basin of attraction of the retrieval solution increases while
the basin of the paramagnetic solution shrinks to zero when
$\varepsilon\to 1$.
This means that the network {\it capacity
diverges} and the {\it retrieval is perfect} for all nonzero
initial states!
This behavior is the analogue of the $\varepsilon_{opt}$
found in the fully connected system where $\varepsilon =
(1+\alpha)^{-1}$ while here we have
\begin{equation}
\varepsilon_{opt} = \frac{1}{1+P/N} \simeq \frac{1}{1+\alpha C/N}
= 1 \;\;\;\; ,
\end{equation}
since $C\ll \ln N$.

For values of $\varepsilon>1$ a rich behavior emerges and the
system  presents a novel kind of retrieval.  The overall behavior
of the system is presented in fig.\ref{pdt0}, showing several
regimes. For a fixed $\varepsilon$ and low values of $\alpha$ the
system is in a retrieval phase (R$_2$) where it oscillates between
a  state with overlap $m$ with the pattern and a state with $-m$
overlap.

As $\alpha$ increases the system enters in a new phase. It starts
oscillating between two distinct values of $m$, say $m_1$ and $m_2$
($|m_1|\neq |m_2|$) while higher periods appear with increasing
values of $\alpha$ leading to a chaotic regime. The route to chaos
followed by the system is not exactly period doubling since it
suffers split
bifurcations \cite{split}:  before doubling the period of the
attractor, the system doubles the number of stable attractors. For
instance, a period 2 cycle splits in two period 2 cycles before
becoming a period 4 one.  The transition from the twofold attractor
to the double
period one occurs when the system enters a superstable orbit (those
that contain the critical points). This
behavior is very similar to that found in the cubic logistic map
$f(m) = \alpha m(1- m^2)$ where only one
split bifurcation is allowed in the periodic region (a system with
two critical points can only have two stable attractors)
\cite{split}. A  representative behavior of the system is shown  in
fig.\ref{ma1}. The basins of each attractor for $\alpha=0.6$ and
$\varepsilon=2$ were also examined: for an initial value of $m$
around 0 the domains are extremely mixed. This behavior is better
understood if one plots the function iterated several times (e.g.,
16) as a function of $m$ (fig. \ref{f16}): the function $f^{(16)}$
is a  set of discontinuous plateau at some values of $m$.

To decide whether a given temporal  sequence is chaotic or not we
calculated the Lyapunov exponent through its definition
\begin{equation}
\lambda \equiv \lim_{n\to\infty} \frac{1}{n} \sum_{i=1}^{n} \ln |
f^{\prime} (m_i) |
\end{equation}
where $n$ is the length of the sequence. In the aperiodic regime
one has $\lambda > 0$ indicating that
the system is chaotic while if its behavior is periodic, $\lambda
< 0$.
It is worth noting that even in
the periodic and in the chaotic attractors, the system always
passes very close to the memorized pattern, as can be observed in
fig.\ref{ma1} for $\varepsilon = 2$. We have verified numerically
that for any point in this
phase (C), the attractors have at least one point with overlap $m$
near 1, hence it is possible to interpret this cyclic or chaotic
phase as an
alternative retrieval phase, in which the system does not
recognize a memory by reaching a fixed point but by wandering
around it.

For large enough $\alpha$ the system enters a
paramagnetic phase (the only fixed point of eq.(\ref{fmt0})) as can
be observed in the phase diagram fig.\ref{pdt0}.
Notice also the similarities between
the $T=0$ phase diagram for the diluted and the fully
connected system: both have a retrieval region around the optimal
value of $\varepsilon$ for all values of $\alpha$ due to the
canceling of the noise term. The
difference is that in the fully connected model $\varepsilon_{opt}=
\varepsilon_{opt} (\alpha)$ while in the diluted one it
is a constant. Also, in the fully connected
case there is not a chaotic phase (not even a periodic
one) because there the connections are
symmetric allowing the introduction of a Lyapunov function (the
Hamiltonian eq.(\ref{tobetruncated})).
We have also determined that, as $\varepsilon\to 1$
\begin{equation}
\alpha_c \sim | 1-\varepsilon |^{-2} \;\;\;\; .
\end{equation}

Figure \ref{pdel1} displays the $T$ versus $\alpha$ phase diagram
for $\varepsilon \leq 1$ where there are only fixed point
attractors:
one with $m=0$ (P) and another with $m\neq 0$ (R$_1$).
In particular, for values of $\varepsilon$
where the transition is continuous the line $T_c$ is independent of
$\varepsilon$ and satisfies
\begin{equation}
\beta_c \int_{-\infty}^{+\infty} {\cal D}y \,
\mbox{sech}^2 \left[ \beta_c y \sqrt{2\alpha}\right] = 1\;\;\;\; .
\end{equation}
If $\varepsilon>1$, the map eq.(\ref{mapm}) presents
a solution with $m = \pm \varepsilon^{1/2}$ at
\begin{equation}
\beta^* = \frac{\sqrt{\varepsilon}}{2} \ln
\frac{\sqrt{\varepsilon}+1}{\sqrt{\varepsilon}-1} \;\;\;\; .
\end{equation}
It means that for $T=T^*=(\beta^*)^{-1}$ the system presents the
above solution for all values of $\alpha$. In the limit
$\varepsilon\to 1$, $T^*=0$ and $m=\pm 1$, as is shown in
fig.\ref{pdt0} and, as $\varepsilon\to\infty$, $T^*\to 1$.
In fig.\ref{pde2} the $T$ versus $\alpha$ phase diagram for
$\varepsilon=2$ is shown. Note that, besides the three phases
appearing at $T=0$, we now found also an R$_1$ phase even for
$\varepsilon>1$. For increasing
values of $\alpha$ the $m=0$ region dominates except
for points around $T^*$. In fig.\ref{e2a2} the behavior of $m$ for
$\varepsilon=2$ and $\alpha=1$ is shown as a function of
$T$ with the corresponding Lyapunov exponent $\lambda$.
Note the backward period bifurcation leading to a retrieval phase
for high temperatures: the system leaves the chaotic regime and
become periodic as the temperature increases until a certain
temperature where it has a fixed point and is able to retrieve the
stored information.

Finally we have analyzed the case in which only two memories have
macroscopic overlap $\kappa$, being the dynamics ruled
by eq.(\ref{dupla}), starting always from the initial condition
$m_{1}(0)=1$ and $m_{2}(0)=\kappa$. The attractors of these
equations were numerically studied and four different phases were
distinguished: a retrieval one (R$_1$) which correspond to a fixed
point with $|m_1| > |m_2| > 0$, another retrieval phase (R$_2$) in
which the system oscillates in a cycle two between the states with
$|m_i | = cte$, $|m_1| > |m_2|$,
a mixed phase (M) with $m_{1} = m_{2} \neq 0$, a
paramagnetic phase (P) with $m_1 = m_2 = 0$ and a cyclic or chaotic
one (C).

For $\alpha=0$, when one considers macroscopic overlap with only
one pattern, the DGZ equation is recovered. But, in the situation
where the system has macroscopic overlap with two memories we do
not recover the DGZ equations, that is, there is still an
$\varepsilon$ dependence on the equations.
Surprisingly, this also holds for $\kappa=0$: the system may
present a cyclic or chaotic regime, absent if one looks for a
retrieval solution. In fig.\ref{k2a0}  we present the phase diagram
$T$ versus $\varepsilon$ for $\kappa = 0.2$ and $\alpha=0$.  For
$T=0$ the system present only two phases: the retrieval one for $
|\varepsilon| <  \kappa^{-1}$ and the cyclic and chaotic one for
$|\varepsilon | > \kappa^{-1}$, while the
mixed phase appears for a given temperature that depends on the
value of $\kappa$. It is important to stress that the boundary
between the retrieval phase and the cyclic or chaotic one for
positive $\varepsilon$ shows the first appearance of cyclic orbits:
inside the C phase there are islands of retrieval.
Also, in the boundary between the mixed phase and the paramagnetic
one, there is a point $\varepsilon^*$, above which the transition
is continuous and discontinuous below. It can be shown that the
transition line in the continuous case is given by $T_c = 1 +
\kappa$.

In the case $\alpha\neq 0$ and $T=0$, the $\alpha$ versus
$\varepsilon$ phase diagram is shown in fig.\ref{k2t0}. This phase
diagram is qualitatively similar to the one where there is overlap
with only one memory (fig.\ref{pdt0}). There is also an value of
$\varepsilon$ where the capacity diverges, but here it is given by
\begin{equation}
\varepsilon_{opt} = 1-\kappa^2 \;\;\; .
\end{equation}
Besides that, the boundary between the paramagnetic phase and the
mixed one is
\begin{equation}
\alpha_c = \frac{2}{\pi} \left( 1 + \kappa \right)^2 \;\;\;.
\end{equation}
In the line $\alpha=0$ of R$_2$ region on the above diagram, there
is always the retrieval solution R$_1$.

\section{CONCLUSIONS}
\label{conc.sec}

We presented an exact solution for the dynamics of a model for
neural networks with high order interactions  that shows a
very rich behavior depending on the parameters $\alpha,
\varepsilon$ and $T$. In analogy with the related fully
connected model, there is an optimal value of $\varepsilon$ for
which the model always retrieves the embedded information (although
the process may not be perfect). It is important to stress that for
some values of $\varepsilon$ the retrieval region is
found at high values of $T$: amazingly, as the temperature is
decreased, the system passes from an disordered phase (P) to an
ordered one (R) and, after passing through a region where it
presents cyclic or chaotic orbits, it reenters in the paramagnetic
region. In this case, the presence of thermal noise may improve the
retrieval abilities of the system.

The periodic or chaotic behavior is only present for
$\varepsilon>1$. Apparently, the reason for this is that below
$\varepsilon=1$ the Hopfield second order term surpass the fourth
order one while for $\varepsilon>1$ it is the reverse. In the later
case, due to the asymmetry on the synapses (the presence of both
$\mu$ and $\nu$ in the learning rule), cycles are introduced. This
is analogous to some models where the task is to retrieve temporal
sequences \cite{kuhn} and the learning rule
explicitly stores the transitions between states. The absence of
this kind of transition in the fully connected model is due to the
possibility of symmetrization of all the connections, and hence
allowing the introduction of an energy function for the system.
Also, if $\varepsilon<1$ the map eq.(\ref{mapm}) is invertible and
continuous, not allowing chaos for one dimensional systems
\cite{ott}.

Due to the very complex nature of the system behavior, it would
also be interesting to study the thermodynamics of the symmetric
diluted case \cite{canning}, trying to see the interplay between
the fully connected case and the highly diluted one. Also, if chaos
can be controlled as proposed by Ott, Grebogi and Yorke \cite{ogy},
the phase boundaries of the above phase diagrams may change,
increasing the retrieval region.

The presence of the mixed (M) phase is an indication that
the study of the generalization abilities of the model may reveal
some interesting features.

At last, a remark on the possible role of chaos on the
problem of perception and recognition is worth. It was
proposed \cite{fre} that the property of large groups of neurons
changing abruptly their activity pattern due to small inputs is
responsible for the flexibility of real nervous
systems to create new patterns when interacting with the outside
world, flexibility that may underlie the learning and creative
processes.

\vskip 2\baselineskip
\noindent
{\large\bf Acknowledgments}: We acknowledge R.M.C. de Almeida for
several interesting discussions and for encouraging this work.
Also, we are in debt with J.C.M. Mombach, F.B. Rizatto,
D. Stariolo and C. Tsallis for useful discussions. F.T.A.
acknowledges the kind hospitality of the IF-UFRGS during his stay
at Porto Alegre. Work partially supported by brazilian agencies
Conselho Nacional de Desenvolvimento Cient\'{\i}fico e
Tecnol\'ogico (CNPq),
Financiadora de Estudos e Projetos (FINEP) and Funda\c{c}\~ao de
Amparo \`a Pesquisa do Estado do Rio Grande do Sul (FAPERGS).

\newpage
\noindent {\Large {\bf Figure Captions}}

\begin{figure}[h]
\caption{Phase diagram $\alpha$ versus $\varepsilon$ at $T=0$ of
the fully connected Truncated model. The lines $\alpha_c^{\pm}$ are
second order while
$\alpha_c'$ is discontinuous. }
\label{fcpdt0}
\end{figure}

\begin{figure}[h]
\caption{Right hand side of eq.(\protect{\ref{fmt0}}) for
$\alpha=0.5$ and two values of $\varepsilon$: 0.5 (dashed) and  2
(solid).}
\label{mapa}
\end{figure}

\begin{figure}[h]
\caption{Fixed points $m$ versus $\alpha$ for several values of
$\varepsilon$. The solid line is for stable solutions while the
dashed is for unstable ones. For $\alpha>2/\pi$ the paramagnetic
solution is always stable.}
\label{meps}
\end{figure}

\begin{figure}[h]
\caption{Phase diagram $\alpha$ versus $\varepsilon$ at $T=0$  for
the diluted Truncated model. The transition is continuous for
$\varepsilon<0.266$. The optimal value of $\varepsilon$ is 1 where
$\alpha_c\to\infty$ as $|1-\varepsilon |^{-2}$. For $\varepsilon>1$
the system can present a periodic or chaotic phase (C) and a
retrieval one (R).}
\label{pdt0}
\end{figure}

\begin{figure}[h]
\caption{Plot of $m$ versus $\alpha$ at $T=0$ for $\varepsilon=2$.
For high values of $\varepsilon$ there is only the
$m=0$ solution while for lower values of  $\varepsilon$ the system
behavior is complex.}
\label{ma1}
\end{figure}

\begin{figure}[h]
\caption{Map eq.(\protect{\ref{mapm}}) iterated 16 times showing
the existence of several plateau for $\varepsilon=2$ and
$\alpha=0.6$.}
\label{f16}
\end{figure}

\begin{figure}[h]
\caption{Phase diagram for some values of $\varepsilon < 1$ showing
the existence of only two phases: a retrieval one ($T<T_c$) and
$m=0$ one ($T>T_c$). For $\varepsilon<0.266$ the line $T_c$ is the
same as of the case $\varepsilon=0$.}
\label{pdel1}
\end{figure}

\begin{figure}[h]
\caption{Phase diagram for $\varepsilon=2$. The retrieval ($R$) and
the periodic or chaotic phases (C) are surrounded by the $m=0$ one
(P). Note that for $T^* \simeq 0.8$ the system
has the solution $m=\pm 1/\protect{\sqrt{2}}$ for all values of
$\alpha$.}
\label{pde2}
\end{figure}

\begin{figure}[h]
\caption{The overlap $m$ versus $T$ for $\varepsilon=2$ and
$\alpha=2$ and the corresponding Lyapunov exponent. The system has
only a retrieval regime at high temperatures.}
\label{e2a2}
\end{figure}

\begin{figure}[h]
\caption{Phase diagram for $\alpha=0$ and $\kappa=0.2$ in the case
where there is macroscopic overlap with two memories (see text).}
\label{k2a0}
\end{figure}

\begin{figure}[h]
\caption{Phase diagram for $T=0$ and $\kappa=0.2$ in the case where
there is macroscopic overlap with two memories (see text). Below
the R$_2$ phase, at $\alpha=0$, the system is in an R$_1$ phase,
and that below M, in a C phase. This compatibilizes this diagram
with the former one (fig.\protect{\ref{k2a0}}).}
\label{k2t0}
\end{figure}


\newpage
\begin{thebibliography}{40}
\def\bi{\bibitem}


\bi{cool} {\sc Coolen A.C.C. and Ruijgrok Th.W.}
1988 Phys. Rev. {\bf A38} 4253 \\ {\sc Kepler T.B. and
Abbott L.F.} 1988 J. Phys. France {\bf 49} 1657 \\
{\sc Gardner E., Derrida B. and Mottishaw P.} 1987
J. Phys. France {\bf 48} 741

\bi{russos} {\sc Patrick A.E. and Zagrebnov V.A.} 1991 J. Phys.
{\bf A24} 3413

\bi{derrida} {\sc Derrida B., Gardner E. and Zippelius A.} 1987
Europhys. Lett. {\bf 4} 167

\bi{amit} {\sc Amit D.J., Gutfreund H. and Sompolinsky H.} 1987
Ann. Phys. NY {\bf 173} 30

\bi{per} {\sc Peretto P. and Niez J.J.} 1986  Biol. Cybern. {\bf
54} 53

\bi{multi} {\sc Gardner E.} 1987 J. Phys. A
{\bf 20} 3453 \\ {\sc Abbott L.F. and Arian Y.} 1987
Phys. Rev. {\bf A36} 5091 \\  {\sc Horn D. and Usher M.} 1988 J.
Phys. France {\bf 49} 389

\bi{kanter} {\sc Kanter I.} 1988 Phys. Rev. {\bf A38} 5972

\bi{tama} {\sc Tamarit F.A., Stariolo D.A. and Curado E.M.F.}  1991
Phys. Rev. {\bf A43} 7083

\bi{wang} {\sc Wang L. and Ross J.} 1991 Phys. Rev. {\bf A44} R2259

\bi{rsana} {\sc de Almeida R.M.C. and Iglesias} 1990 Phys. Lett. A
{\bf 146} 239 \\
{\sc Arenzon J.J., de Almeida R.M.C. and Iglesias J.R.}
1992 J. Stat. Phys. {\bf 69} 385

\bi{trs} {\sc Arenzon J.J., de Almeida R.M.C., Iglesias J.R., Penna
T.J.P. and de Oliveira P.M.C.} 1993 Physica A {\bf 197} 1 \\ {\sc
Arenzon J.J. and R.M.C. de Almeida} 1993 Phys. Rev. {\bf E48} 4060

\bi{hop} {\sc Hopfield J.J.} 1982  Proc. Natl. Acad. Sci. USA {\bf
79} 2554

\bi{model} {\sc Hertz J., Krogh A. and Palmer R.G.} 1991
Introduction to the Theory of Neural Computation
(Addison-Wesley Publishing Company)

\bi{split} {\sc Testa J. and Held G.A.} 1983 Phys. Rev. {\bf A28}
3085

\bi{fre} {\sc Skarda C.A. and Freeman W.J.} 1987 Behavioral and
Brain Sciences {\bf 10} 161

\bi{kuhn} For a review see {\sc Kuhn R. and van Hemmen J.L.} in
``Physics of Neural Networks'' (E. Domany, J.L. van Hemmen and K.
Schulten eds.), Springer 1990.

\bi{canning} {\sc Canning A. and Gardner E.} 1988 J. Phys. {\bf
A21} 3275

\bi{ogy} {\sc Ott E., Grebogi C. and Yorke J.} 1990 Phys. Rev.
Lett. {\bf 64} 1196

\bi{ott} {\sc Ott E.} 1993 Chaos in Dynamical Systems (Cambridge
University Press)

\end{thebibliography}
\end{document}